# Development of an ASD IC for the Micro Pixel Chamber

R.Orito, O.Sasaki, H.Kubo, K.Miuchi, T.Nagayoshi, Y.Okada, A.Takada, A.Takeda, T.Tanimori, and M.Ueno

*Abstract*— A new amplifier-shaper-discriminator (ASD) chip was designed and manufactured for the Micro Pixel Chamber ($\mu$-PIC). The design of this ASD IC is based on the ASD IC (TGC-ASD) for the Thin Gap Chamber in the LHC Atlas Experiment. The decay time constant of the preamplifier is 5-times longer than that of the TGC-ASD, and some other modifications have been made in order to improve the signal-to-noise ratio of the $\mu$-PIC. The ASD IC uses SONY Analog Master Slice bipolar technology. The IC contains 4 channels in a QFP48 package. The decay time constant of the preamplifier is 80 ns and its gain is approximately 0.8 V/pC. The output from the preamplifier is received by a shaper (main-amplifier) with a gain of 7. A baseline restoration circuit is incorporated in the main-amplifier, and the current used for the baseline restoration is 5-times smaller than that of the TGC-ASD. The threshold voltage for the discriminator section is common to the 4 channels and their digital output level is LVDS-compatible. The ASD IC also has an analog output of the preamplifier. The equivalent noise charge at the input capacitance of 50 pF is around 2000 electrons. The power dissipation with LVDS outputs (100 $\Omega$ load) is 57 mW/ch. Using this ASD, the analog output voltage from the signal of the $\mu$-PIC is about 2-times higher than the case of using the TGC-ASD. As a consequence, the MIPs tracking performance of the Time Projection Chamber (TPC) with the $\mu$-PIC was improved. The performance of the ASD IC and an improved tracking performance of the TPC are reported.

## I. INTRODUCTION

The Micro Pixel Chamber ($\mu$-PIC)[1][2], which is a micro-pattern gaseous detector based on double-sided printed circuit board (PCB) technology, has been developed for X-ray, gamma-ray and charged particle imagings. In the $\mu$-PIC, anode and cathode strips are formed orthogonally on both sides of a polyimide substrate with a pitch of 400 $\mu$m, as shown in Fig. 1. This micro-structure makes it possible to track particles more precisely than by using conventional wire chambers. Furthermore, the $\mu$-PIC has a pixel-type structure and a thick substrate, therefore a stable operation is realized at a higher gain than that of the MSGCs[3]. Using the $\mu$-PIC with a $10 \times 10$ cm$^2$ detection area[4], clear two dimensional X-ray images were obtained[5][6]. Using the $\mu$-PIC, the micro Time Projection Chamber ($\mu$-TPC) was also made for particle tracking[7], and low-energy proton tracks were clearly detected in an Ar-C$_2$H$_6$ gas mixture at 1 atm[8][9]. Intermediate gas multipliers, such as GEM[10] or a capillary plate[11], are not needed for detecting high dE/dx particles. This simple structure is a great advantage of the $\mu$-TPC. Presently the remaining problem of the $\mu$-TPC is low efficiency for low dE/dx particles or Minimum Ionizing Particles (MIPs). A narrow spacing of the electrodes realizes fine particle tracking, but it produced fewer electrons per electrode than that of ordinary wire chambers. In micro-pattern detectors, it is not an easy task to raise the gas gain, because the electric field is very intensive around a narrow space between the electrodes. A detailed study of the electrode structure[12][13] and a technique of manufacturing the electrodes accurately, as designed, are necessary for improving the gas gain. Therefore, to increase the gain, optimizing the amplifier of the chamber is much easier than changing the electrode structure. At first, we used the TGC-ASD[14] in which the decay time constant of the preamplifier was 16 ns, for a readout of the $\mu$-PIC. However, this decay time constant was too short to collect all pulse charges from the $\mu$-PIC. Therefore, one way to improve the efficiency of $\mu$-PIC for MIPs is to develop an amplifier that has a sufficient decay time constant of the preamplifier to collect all charges from the $\mu$-PIC. Based on a simulation, when we use a preamplifier with a 80 ns integtration constant for readout of the $\mu$-PIC, the signal from the $\mu$-PIC is 2-times higher than in the case of using the preamplifier with a 16 ns decay time constant. Therefore, we have developed an ASD IC that includes the preamplifier with an 80 ns decay time constant ($\mu$PIC-ASD).

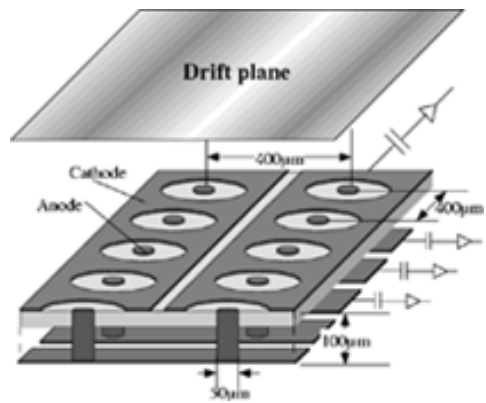

Fig. 1. Schematic structure of the $\mu$-PIC.

## II. CIRCUIT DESIGN

### A. Technology

Because of the thin spacing of the electrodes, the detector capacitance of the $\mu$-PIC is relatively large. The measured capacitance of the $\mu$-PIC is about $1.5 \times 10^{-1}$ pF per pixel, 38 pF per 10 cm length strip. For the $\mu$-PIC with an area of over $30 \times 30$ cm$^2$, low-noise amplifier is needed. Therefore, transistors with a large $g_m$ are preferred[15]. Hence, we chose to base the amplifiers on bipolar transistors. A chip has been developed at SONY corporation, using their bipolar 'Analog Master Slice Process'. This semi-custom process provides pre-fabricated NPN and PNP transistors, resistors and capacitors, so that a designer has to design using these elements that are pre-determined beforehand for the silicon wafer. The base structure that we used contains 850 NPN transistors, 384 PNP transistors, 1738 resistors and 42 capacitors, totaling approximately 1000 usable elements. The standard transistor has $f_T$ = 3.2 GHz. The low-noise transistor has $f_T$ = 950 MHz and base-spread resistance $r_{bb'}$ =17.5 $\Omega$. There are also PNP type transistors, of which the standard one has $f_T$ = 300 MHz. The capacitors are of 2 pF and 20 pF value, totaling 408 pF (Metal Insulator Semiconductor: MIS capacitor). The resistors are of either 8 k$\Omega$ or 2.5 k$\Omega$ (poly-silicon), 297 $\Omega$ and 129 $\Omega$ (diffused). These are used in combination.

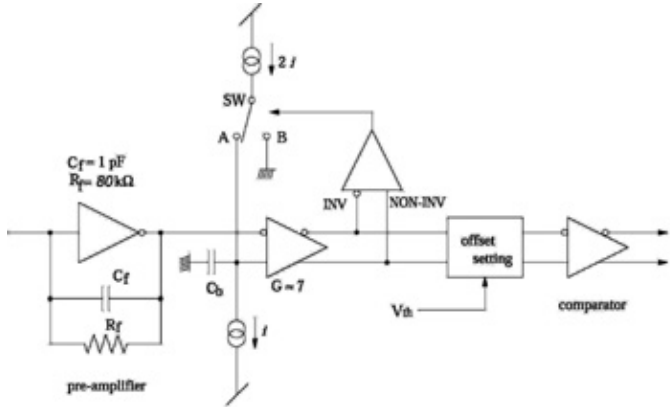

Fig. 2. Block diagram of the ASD chip.

### B. Circuit Diagram

A block diagram of the ASD chip is shown in Fig. 2, with schematics given in Figs. 3 and 4. The circuit is based on the TGC-ASD with some modifications for the $\mu$-PIC. The first stage is a common-emitter cascode charge amplifier. The input stage of the preamplifier is implemented with the low-noise NPN transistor with an $r_{bb'}$ of 17.5 $\Omega$. The relatively large capacitance (higher than 10 pF between the collector and substrate) of the transistor disfavors the use of the transistor in a common-base configuration, which is usually employed in preamplifiers for chambers. The collector current of the head transistor is set high (0.4 mA) so as to achieve a large $g_m$, which has an advantage to achieve lower noise at a larger detector capacitance. The decay time constant of the preamplifier is set at to 80 ns. The gain of the preamplifier stage is approximately 0.8 V/pC. An emitter follower output of this stage is provided for monitoring. The second stage consists of the main-amplifier with a baseline restorer and differential outputs. A gain of the main-amplifier is about 7. Depending on the output differential signal level seen by the switch control section, the switch connects to the "A" side or "B" side of Fig. 2. When the switch is connected to the "A" side, the capacitance $C_b$ will be charged from the current source by an amount of "i". When the switch is connected to the "B" side, the capacitor will be discharged by an amount "i", resulting in stabilized DC output levels, or baseline restoration. The current of "i" is restricted to being 5-times smaller than that of the TGC-ASD, because a large current reduces the pulse height fed to the comparator. Following the main-amplifier, there is an offset setting that transforms the main-amplifier output levels required at the inputs to the comparator, where the offset voltage is controlled by a DC voltage ($V_{th}$) supplied from outside of the chip. A comparator circuit is shown in Fig. 4. Its outputs conform to the Low Voltage Differential Signalling Standard (LVDS), to minimize power and to assure drivability and immunity against a noise. By design, this circuit can be used for both anode and cathode strips signals by setting an appropriate threshold level. Table I gives a summary of this chip's characteristics.

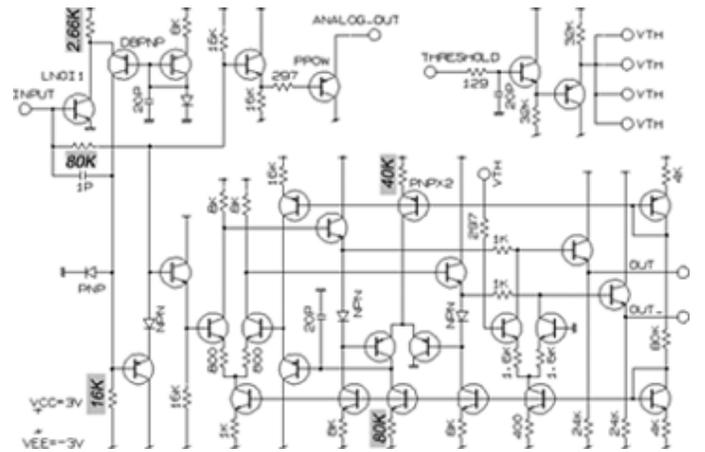

Fig. 3. Schematic of the preamplifier, baseline restorer and main-amplifier. The hatched areas represent modifications from the TGC-ASD.

### C. Simulation

Fig. 5 shows the result of a PSPICE simulation of the preamplifier output, the main-amplifier differential outputs and the comparator LVDS outputs against impulse inputs of -0.1 $\sim$ -0.5 pC charge. The dynamic range of the preamplifier is from -1.4 to 2 pC for negative/positive impulse charge inputs.

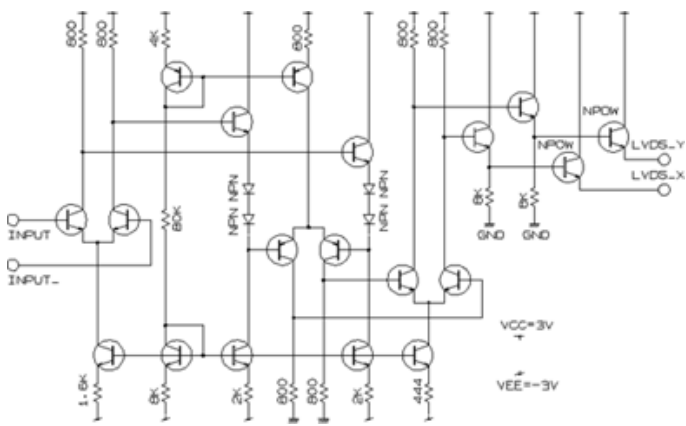

Fig. 4. Schematic of the comparator circuit.

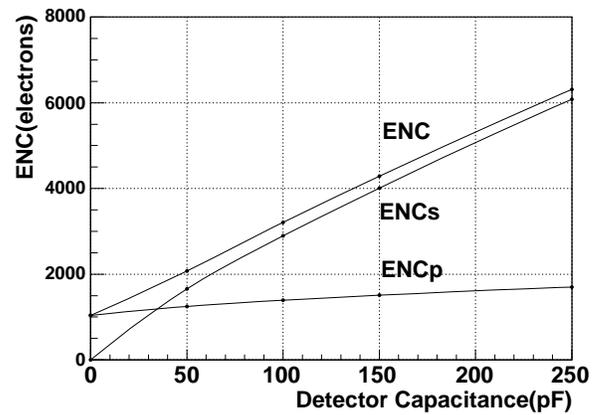

Fig. 6. Calculated ENC as a function of the detector capacitance.

The dynamic range and the gain of the preamplifier observed at the buffered direct output depend on the external load, and are less than the internal one. The rise time of the preamplifier output is 8 ns. The threshold level can be controlled between -0.1 to 0.1 pC for impulse inputs. The circuit can successfully accept signals of 1MHz or higher frequency. Fig. 6 shows the calculated equivalent-noise-charge (ENC) as a function of the detector capacitance. At 50 pF input capacitance, ENC is around 2000 electrons r.m.s..

TABLE I
ASD CHIP CHARACTERISTICS.

| Process | Sony Analog Master Slice Process bipolar, semi-custom |
|---|---|
| Specification | preamplifier with a gain of 0.8 V/pC 80 ns decay time constant input impedance of around 370 $\Omega$ open emitter analog outputs main-amplifier with a gain of 7 baseline restoration circuits comparator with LVDS outputs ENC $\sim$ 2000 electrons at $C_D$ = 50 pF 4 channels in a QFP48 plastic package threshold voltage : common for all 4 channels required voltage +/- 3V, GND 56.7mW/ch when driving a 100 $\Omega$ load (+3V:15.8mA -3V:3.11mA 42.7mW in ASD chip and 14 mW at LVDS receiver end ) |

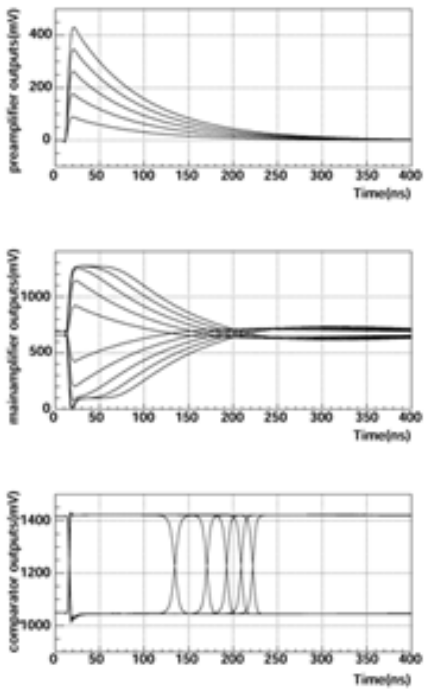

Fig. 5. PSPICE simulations of the preamplifier, main-amplifier and comparator outputs for -0.1 to -0.5 pC impulse inputs.

### D. IC Layout

4 channels of ASDs were fabricated on a 3.1 mm × 3.1 mm die, as shown in Fig. 7. The threshold voltage is common to 4 channels. In the layout work of the IC, we paid much attention to reduce the interference between the analog and digital signals and crosstalk among the channels. Both the ground and power patterns and I/O pads for the analog parts are separated from those for the digital parts. This chip is housed in a QFP48 package. The pins of the package were assigned while keeping right-left symmetry. For protection from static charge, diodes are attached between all of the I/O pad and the most positive/negative voltage except those for the ground and DC powers.

## III. PERFORMANCE

### A. ASD Performance

The analog and digital signals from the ASD chip for impulse inputs from 0.1 to 0.4 pC are shown in Fig. 8. In this figure, the open-emitter is pulled up to 3 V by 510 $\Omega$ and loaded to 50 $\Omega$. The common threshold is set to -50 mV. The measured analog output of the preamplifier is shown in Fig. 9, for the case when the open-emitter is pulled up to 3 V by 510 $\Omega$ and loaded to 1 k$\Omega$. The negative-positive symmetry of the linearity was improved from that of the TGC-ASD. The ENC was measured as a function of the input capacitance, as shown in Fig. 10. We calculated the ENC using design parameters of the preamplifier and measured impulse response of the evaluation system. The calculation reproduces the measured data with an $r_{bb'}$ of 20 $\Omega$ and an $h_{fe}$ of 90. The ENC was reduced from the TGC-ASD, because of a longer decay time constant of the preamplifier. The crosstalk among channels was less than 0.5% when analog outputs were left open. If the open-emitter buffer for the analog output drove a 50 $\Omega$ load, the crosstalk became 3-times larger. Those performances are of pre-production samples (240 pieces). Since the design of the ASD is simple, this ASD can be used not only for the $\mu$-PIC, but also for general gas-chambers. Good radiation hardness is also expected[16]. Therefore, for requests to use for various gas-chambers in different experiments, mass-production (24 k pieces) of this ASD was started in September, 2003.

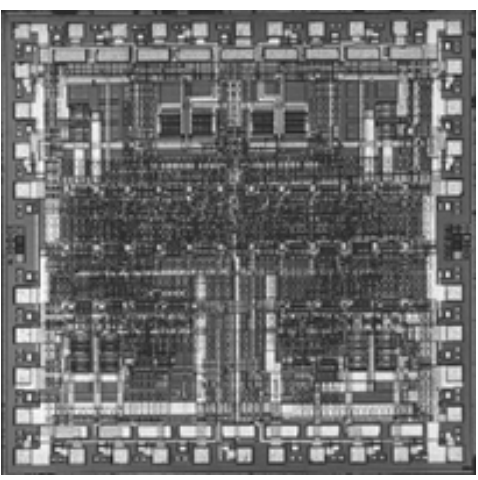

Fig. 7. Micro-photograph of the ASD chip.

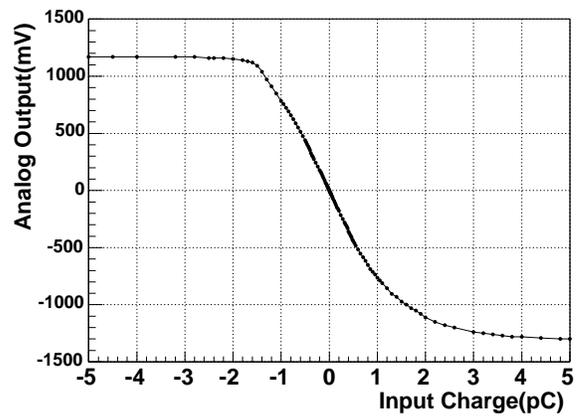

Fig. 9. Measured analog output as a function of the input charge.

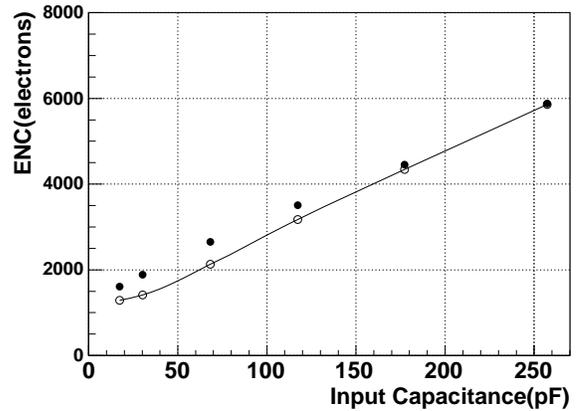

Fig. 10. Measured ENC(filled circle) and calculated ENC(open circle) using the design parameters of the preamplifier and the measured impulse of the evaluation system.

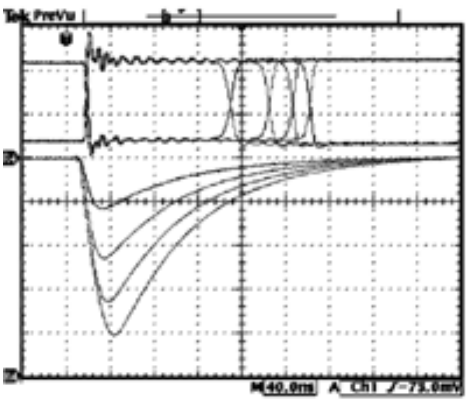

Fig. 8. Analog and digital signals from the ASD for impulse inputs from 0.1 to 0.4 pC. The voltage and time axes are 50 mV/div and 40 ns/div, respectively.

### B. Improvement of the $\mu$-TPC performance

We measured signals from the 10×10 cm$^2$ $\mu$-PIC using this ASDs. All analog outputs from anode strips were summed. All measurements were performed under the condition of Ar-$C_2H_6$(8:2) gas at 1 atm pressure. When the detector was irradiated with 5.9 keV X-rays from an $^{55}$Fe source, the pulse height of the analog outputs was 2-times higher than that of the TGC-ASD, as shown in Fig. 11. The energy spectra of $^{55}$Fe source taken by the flash ADC (FADC) are shown in Fig. 12. The energy resolutions (FWHM) with 10×10 cm$^2$ area of the $\mu$-PIC were 30.4% ($\mu$PIC-ASD) and 43.5% (TGC-ASD). Fig. 13 shows an example of three-dimensional tracks of cosmic-ray muons. The efficiency for cosmic-ray muons using $\mu$PIC-ASD was about 20% with a gas gain of ∼3000. This is a much higher efficiency than in the case of using the TGC-ASD.

We are developing a Compton gamma-ray imaging detector with the gaseous $\mu$-TPC and a scintillation camera[17][18]. In this detector, we need to detect fine tracks of the keV-MeV electrons. Using the $\mu$PIC-ASD, the efficiency of the electron track was also improved. Fig. 14 shows the track of a keV-MeV Compton-scattered electron taken by the $\mu$-TPC. The efficiency for low dE/dx particles or MIPs will be almost 100% when the new $\mu$-PIC[6] is operated stably. The results of efficiency tests will be reported in detail in another paper.

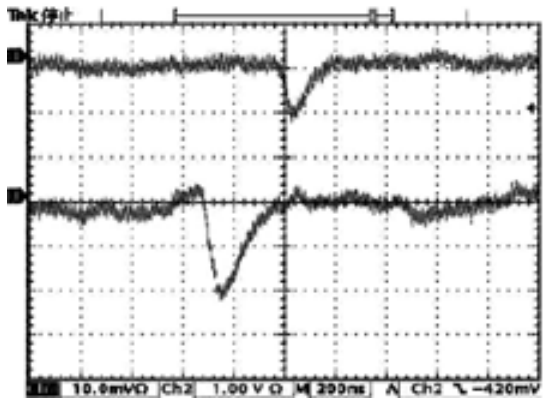

Fig. 11. Signals from the $\mu$-PIC. TGC-ASD (upper) and $\mu$PIC-ASD (lower), respectively.

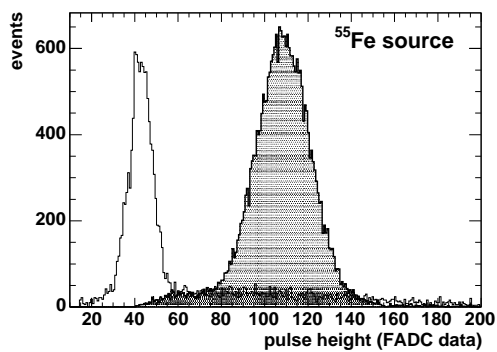

Fig. 12. $^{55}$Fe spectrum of the $\mu$-PIC. The filled histogram and the unfilled histogram represents cases using the $\mu$PIC-ASD and the TGC-ASD, respectively.

## IV. SUMMARY

We have developed a new Amplifier-Shaper-Discriminator IC for the Micro Pixel Chamber. The technology involved the SONY semi-custom Analog Mater Slice bipolar process, and 1 chip contains 4 channel ASDs. The decay time constant of the preamplifier is 80 ns. Four channel analog outputs and digital outputs are equipped. The produced pre-production samples (240 pieces) performed to the specifications. Using this ASD, the tracking performance of $\mu$-TPC for MIPs was improved. The mass production of the chips was started in September, 2003.

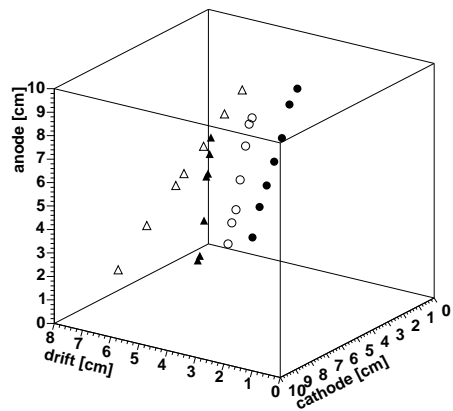

Fig. 13. Three-dimensional tracks of cosmic-ray muons.

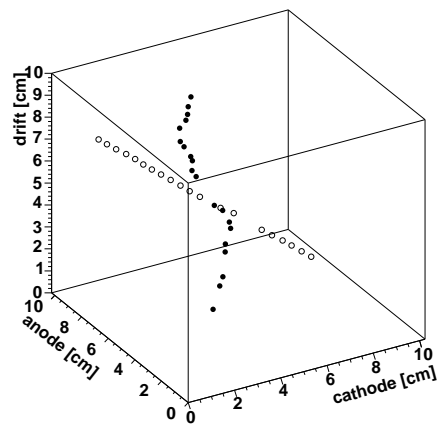

Fig. 14. Three-dimensional tracks of a Compton-scattered electron from a $^{137}$Cs source (filled circle) and an 0.8 GeV proton (open circle).

## ACKNOWLEDGMENTS

We would like to acknowledge Mr. K.Sinozaki and Mr. M.Nakamura of SONY Corporation for their support in IC production. We also thank Mr. M.Ikeno of KEK for his support in the IC test. This work is supported by a Grant-in-Aid for the 21st Century COE "Center for Diversity and Universality in Physics", a Grant-in-Aid in Scientific Research of the Japan Ministry of Education, Culture, Science, Sports and Technology, and "Ground Research Announcement for Space Utilization" promoted by Japan Space Forum.